\documentclass{ws-p8-50x6-00}

\begin{document}

\title{An Approximate Light Cone Method to Investigate Meson Structure}

\author{Sudip K. Seal and Matthias Burkardt}

\address{Department of Physics, New Mexico State University,
\\ Las Cruces, NM 88011, U.S.A.\\
E-mail: skseal@dunkel.nmsu.edu, burkardt@dunkel.nmsu.edu}


\maketitle

\abstracts
{We present results obtained using a transverse lattice formulation of QCD
to calculate light and heavy meson observables.}

\section{Introduction}
It is well-established that the light-front formulation of QCD 
make calculations of observables in high-energy phenomena like
DIS considerably easier while providing a more natural 
interpretation of the same \cite {bpp}. Transverse lattice 
formulation of QCD attempts to exploit the benefits \cite{bardeen,mb:adv,dalley:aip} 
from using the LF gauge and the well-known advantages of the more 
conventional Euclidean and Hamiltonian lattice formulations. 

Transverse lattice techniques were employed quite successfully to 
pure glue calculations \cite{mb:bob,brett}. Encouraged by these results,
the {\em femtoworm approximation}, which we describe briefly in the next section,
was conceived to include the fermions on the transverse lattice, as well. 
\cite{mb:hala,meson:sd,meson:us,heavy:us,strange:us}

In this paper, we briefly describe the femtoworm approximation in section 2, 
referring the reader to relevant references for the most part. Some typical results
and conclusions are included in section 3. 

\section{The Hamiltonian of the Femtoworm Approximation}
Using light cone gauge requires that the longitudinal light cone space,
$x^+$  and light cone time, $x^-$ variables be 
redefined in terms of the equal time longitudinal space, $x^0$ and time, $x^3$ coordinates 
as 
$x^\pm=x^0\pm x^3$. The transverse directions are the same as in equal time definitions but 
discretized. This enables the use of a gauge invariant regulator in the theory while 
simultaneously retaining boost invariance in the longitudinal direction. 
In addition, the femtoworm approximation is based on the following considerations:
\newline
$\bullet$ We work in the large $N_c$ limit.
\newline
$\bullet$ A light cone Tamm-Dancoff truncation is applied so that the Fock state expansion of the
meson states include only 2-particle and 3-particle Fock components. 
\newline 
$\bullet$ One link approximation: 
The maximum number of link fields that can separate a $q$ and an $\bar{q}$ is one.
\newline
$\bullet$ Interactions between 2-particle to 3-particle states (and vice versa) can take place
only through two interaction terms of the Hamiltonian -- one that flips the spin of the 
interacting fermions and the other that does not. 

The 2-particle states consist of a $q\bar{q}$ pair sitting on the same $\perp$ lattice site.
The 3-particle states consist of a $q$ and an $\bar{q}$ occupying neighboring sites 
separated by a link field (the gauge degrees of freedom in this approach).
The 2-sector part of the Hamiltonian contains two terms that describe the kinetic energies
of the $q$ and the $\bar{q}$, respectively. It also contains a 
longitudinally confining Coulomb interaction term acting between the pair. 
The 3-sector part of the Hamiltonian contains three terms that describe the kinetic energies
of the $q$, the $\bar{q}$ and the link field, respectively. There 
are two more terms -- one describes a longitudinally confining
 Coulomb interaction between the $q$ and
the link field at the site where the $q$ resides and another that describes a 
longitudinally confining
 Coulomb interaction between the $\bar{q}$ and the link field at the neighboring
site where the $\bar{q}$ resides. Note that there are no Coulomb terms connecting the $q$ and the
$\bar{q}$ on neighboring sites. This is because these interaction terms are local on the 
$\perp$ lattice sites. Further, such interactions would yield non-planar diagrams which 
would violate our large $N_c$ approximation.

Transverse propagation of the mesons take place through the two interaction terms that 
connect the two Fock states. The meson propagates by emitting (2- to 3-sector) or absorbing 
(3- to 2-sector) a link field. When a meson is in the 2-sector (i.e., $q$ and $\bar{q}$ are 
sitting on the same $\perp$ lattice site), one of it's constituents
can hop to a neighboring site emitting a link field either through the
spin flip term in which case it's spin will flip or through the spin non-flip
term in which case it's spin will not flip. The meson in this configuration (a $q$ and an 
$\bar{q}$ on neighboring sites separated by one link field) 
represents it's 3-particle state. When the meson is in this state, one of the fermion
degrees of freedom hops and joins the other fermion on a neighboring site by absorbing
 the link field either through the spin flip in which case 
it's spin will flip or through the spin non-flip interaction term in which case it's spin will 
not flip. The nature of the propagation is reminiscent of that of an inchworm on a scale of
a femtometer which motivated the name {\it femtoworm}.

The strength of spin flip and spin non-flip interaction terms are labeled $m_v$ and $m_r$, 
respectively. Though, in general, a link field emitted through a spin flip (spin non-flip) 
interaction term can be absorbed through a spin non-flip (spin flip) interaction term, such an 
interference term is disallowed in the ${\bf k_\perp=0}$ case. This and other spin dynamics 
are discussed in details in \cite{mb:hala,meson:us}.

The parameter space that defines the above $q\bar{q}$ bound state Hamiltonian 
contain the following parameters:
\newline
$\bullet$ kinetic masses for the light quark in the 2 and 3 particle Fock sector.
\newline
$\bullet$ kinetic masses for the strange quark in the 2 and 3 particle Fock sector.
\newline
$\bullet$ kinetic mass for the link field.
\newline
$\bullet$ spin flip and non-flip hopping terms for the light quark.
\newline
$\bullet$ spin flip and non-flip hopping terms for the strange quark and
\newline
$\bullet$ longitudinal gauge coupling.

As explained in \cite{meson:us}, this parameter space is constrained by: 
\newline
$\bullet$ {\bf imposing Lorentz symmetry.} 
This is done by demanding that the eigenstates of the Hamiltonian have the same 
dispersion relation.
\newline
$\bullet$ {\bf imposing chiral symmetry in the spontaneously broken form.} 
This is implemented by tuning the spin flip and spin non-slip hopping terms to 
yield the observed $\pi-\rho$ mass splitting and
\newline
$\bullet$ {\bf fixing the physical string tension.} 
We used the string tension as obtained from 
glueball calculations \cite{brett}.  

\section{Results and Conclusions}
We have studied the LF momentum distribution amplitudes of several light mesons ($\pi$, $\rho$),
mesons with strange quarks ($KK^*$ system, $\phi$) as well as heavy mesons ($B$-meson).
For each of the above, we have also calculated their respective parton distribution functions. 

The LF momentum distribution of the pion [Fig.1(a)] is quite close to the asymptotic shape.
The first five even coefficients 
of a Gegenbauer projection of the momentum distribution, 
$\psi(x)=4x(1-x)\sum_{i=0}^{\infty}a_iG^{\frac{3}{2}}_i(1-2x)$, are $\{1.5,\;0.102,\;-0.006,\;
-0.005,\;-0.001\}$.  
The tail of the parton distribution function [Fig.1(b)] is somewhat higher than what is 
observed. 
Because there is only one link, most of the 
longitudinal momenta  close to $x=1$ are carried by the fermions.

The dashed line in Fig.1(c) shows the $\sqrt{z}$ behavior of the $B$-meson LF momentum 
distribution in the heavy quark limit. For phenomenological applications \cite{aa}
it is useful to have numerical estimates for the
moments \cite{heavy:us} 
(the normalization is $\int_0^\infty dz \phi_\infty(z) dz =1$) for which we find
$\int dz z \phi_\infty(z)=1.51\,GeV$ and
$\int dzz^{-1} \phi_\infty(z)=1.22\,(GeV)^{-1}.$
For the decay constant \cite{heavy:us}, 
which plays also an important role in mixing phenomenology, we find
$f_B\approx 240 MeV\pm 20 MeV$. Fig. 2 shows the Isgur-Wise form factor of the $B$-meson
in the heavy quark limit which allows one to extract the $V_{cb}$ element of the CKM matrix 
from the zero recoil point of a semi-leptonic decay channels like  
$B\rightarrow \bar{D}^*l \nu$.
We obtain values of $V_{cb}$ in the range $0.0415-0.0420$ which is very consistent with the 
experimental data \cite{exp}.

Since we used basis function techniques and extrapolated the 
results in Hilbert space dimensions, uncertainties from Hilbert space truncation are under 
control. Systematic errors arising from the Fock space truncation cannot be estimated 
unless higher Fock components are included. Thus, a main improvement over this work would be 
to include such Fock components. For example, we 
expect that including such components will significantly reduce the 2-particle Fock space
amplitude, thereby reducing the decay constants which are calculated \cite{meson:us,heavy:us}
from the normalizations of the LF distribution amplitudes. Other observables are expected to 
be effected in a similar manner.

\begin{figure}
\unitlength0.9cm
\begin{picture}(15,7.0)(-14.0,4.0)
\includegraphics{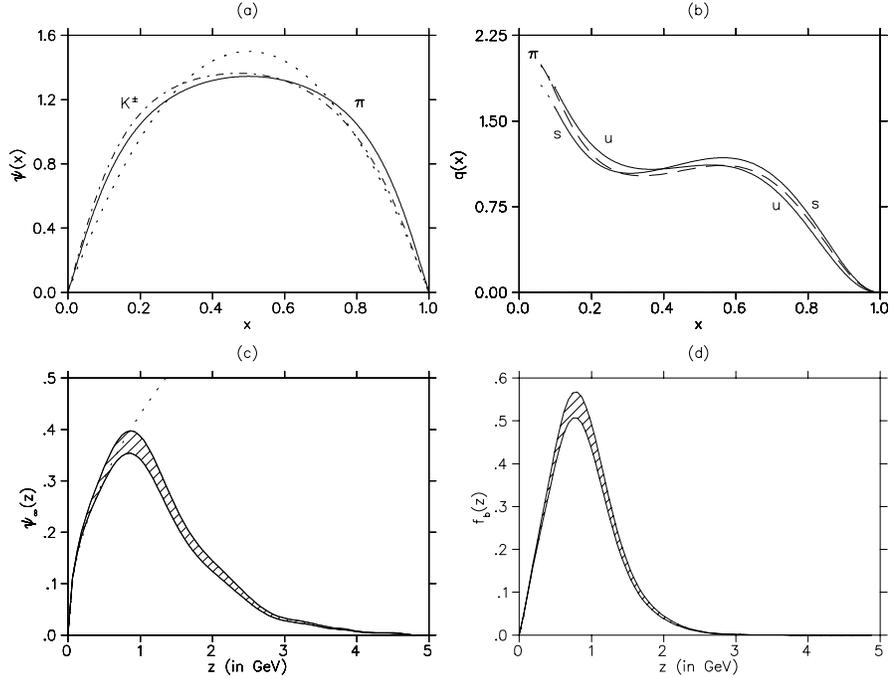}
\end{picture}
\vspace{1.0in}
\caption{(a) LF momentum distribution amplitude of the $\pi$ (bold), $K^{\pm}$ (dashed) and the
asymptotic shape (dotted). (b) Parton distribution function of the $u$-quark in $\pi$ (dashed),
the $u$ and $s$ distributions in $K^{\pm}$. (c) $B$-meson distribution amplitude in the heavy 
quark limit. (d) $B$-meson LF momentum distribution in the heavy quark limit. Systematic 
uncertainties arising from the numerical extrapolation are indicated by the shaded area.}
\label{fig:form}
\end{figure}
\begin{figure}
\unitlength0.9cm
\begin{picture}(15,7.0)(-11.0,0.0)
\includegraphics{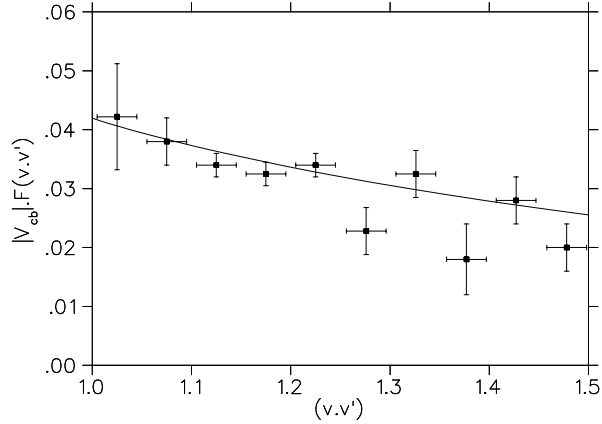}
\end{picture}
\vspace{-0.5in}
\caption{Isgur-Wise form factor in the heavy quark limit.
Uncertainties from the $M_b\rightarrow \infty$ limit were very small for this 
observable. The experimental data is from Ref. [12] .}
\label{fig:form}
\end{figure}

For the heavy quark observables, $1/M_b$  and perturbative
corrections are expected to contribute 
at zero recoil ($v.v'=0$). We were not able to estimate the former because we
worked in the $M_b\rightarrow\infty$ limit. Also, it is not known how to reach 
small lattice spacings on the $\perp$ lattice and as such an estimate of the latter 
corrections are beyond the scope of this technique. Despite this caveat, we feel that 
with further improvements, this technique holds promise.

\end{document}